\documentclass[aps,prx,twocolumn,superscriptaddress,showpacs]{revtex4-1}
\usepackage{amsmath,amssymb,graphics,epsfig,epstopdf,color,verbatim,ulem,braket,tabularx}
\usepackage{multirow}
\usepackage[colorlinks,linkcolor=blue,citecolor=blue,urlcolor=blue,bookmarks=false]{hyperref}

\usepackage{listings}
\usepackage{cancel}
\usepackage{soul}

\begin{document}

\title{Entropy and electronic orders of the three-orbital Hubbard model with antiferromagnetic Hund coupling} 

\author{Changming Yue}
\email{changming.yue@unifr.ch}
\affiliation{Department of Physics, University of Fribourg, 1700 Fribourg, Switzerland}

\author{Shintaro Hoshino}
\affiliation{Department of Physics, Saitama University, Saitama 338-8570, Japan}

\author{Philipp Werner}
\email{philipp.werner@unifr.ch}
\affiliation{Department of Physics, University of Fribourg, 1700 Fribourg, Switzerland}

\begin{abstract}
An antiferromagnetic Hund coupling in multiorbital Hubbard systems induces orbital freezing and an associated superconducting instability, as well as unique composite orders in the case of an odd number of orbitals. While the rich phase diagram of the half-filled three-orbital model has recently been explored in detail, the properties of the doped system remain to be clarified. Here, we complement the previous studies by computing the entropy of the half-filled model, which exhibits an increase in the orbital-frozen region, and a suppression in the composite ordered phase. The doping dependent phase diagram shows that the composite ordered state can be stabilized in the doped Mott regime, if conventional electronic orders are suppressed by frustration. While antiferro orbital order dominates the filling range $2\lesssim n \le 3$, and ferro orbital order the strongly interacting region for $1\lesssim n < 2$, we find superconductivity with a remarkably high $T_c$ around $n=1.5$ (quarter filling). Also in the doped system, there is a close connection between the orbital freezing crossover and superconductivity. 
\end{abstract}

\maketitle

\section{Introduction}

The ferromagnetic Hund coupling in conventional multi-orbital Hubbard models leads to spin-freezing and bad metal behavior in a wide filling, interaction, and temperature range \cite{prl2008_Werner_spinfreezing,PhysRevLett.107.256401}. This physics is crucial for understanding the normal-state properties of an interesting class of correlated materials, the so-called Hund metals \cite{Haule_2009,YinZP_2011_NatMatt,annurev_Hund_2013}, and it is closely linked to unconventional superconductivity \cite{Hoshino2015,Werner2016}. Related phenomena appear in multi-orbital Hubbard models with negative (antiferromagnetic) Hund coupling \cite{Erio_prb1997_JahnTeller,Nomurae1500568,Steiner2016}, where the physics turns out to be even more complex because of the appearance of odd-frequency orbital orders \cite{Hoshino2017}. Particularly interesting is the case of the three-orbital Hubbard model with antiferromagnetic Hund coupling, which is relevant for the description of fulleride compounds \cite{RevModPhys.69.575,Capone2364,Gunnarsson_Fullerides_Book_2004,RevModPhys.81.943,Nomura2012}. Several recent studies \cite{Hoshino2017,Ishigaki2018,Ishigaki2019} have considered the half-filled case and revealed (i) a connection between the unconventional $s$-wave superconductivity and an orbital-freezing phenomenon, (ii) the existence of spontaneous orbital selective Mott (SOSM) phases, which can be interpreted as odd-frequency orbital orderings, in the vicinity of the Mott transition, (iii) the existence of a spontaneous orbital selective superconducting phase, and (iv) various types of staggered or uniform orbital and magnetic orders. These results provide a general understanding of the phase diagram of alkali-doped fullerides (K$_3$C$_{60}$), with three electrons in three $t_{1u}$ molecular orbitals, including the recently reported Jahn-Teller metal phase \cite{Zadik2015}, which can be identified with the SOSM phase \cite{Hoshino2017}. 

Last year, the first single crystals (thin films) of fulleride compounds were grown epitaxially, and this experimental technique allows to control the filling over a wide range via doping \cite{Ren2020,Han2020}. It is thus interesting to extend the theoretical studies of the three-orbital Hubbard model with negative Hund coupling to the doped regime. While a previous work has reported some results for a doped realistic model of fulleride superconductors \cite{Misawa2017}, we currently lack a complete picture of how the doping affects the orbital freezing, superconductivity, SOSM states and other electronically ordered phases.  

In this paper, we try to fill this gap by presenting a systematic study of the crossovers and phase transitions in the half-filled and doped three-orbital Hubbard model with degenerate bands. After introducing the model in Sec.~\ref{sec:model} we start by revisiting the half-filled phase diagram and discuss in Sec.~\ref{sec:entropy} the effect of the orbital-freezing crossover, SOSM and Mott transitions on the entropy. We then turn our attention to electronic orders and show in Secs.~\ref{sec:doped_mott} and \ref{sec:doped_sosm} how the orbital-ordered, charge-ordered and (metastable) SOSM phases depend on the filling. In Sec.~\ref{sec:susc}, we clarify the connection between superconductivity and orbital freezing in the doped system. Section~\ref{sec:summary} contains a summary and conclusions. 

\section{Model and Method}
\label{sec:model}

We use dynamical mean field theory (DMFT) \cite{Hartmann_1989a,Hartmann_1989b,PhysRevLett.62.324,rmp_68_13_dmft_1996,rmp_dftdmft} to simulate the three-orbital Hubbard model on an infinitely connected Bethe lattice. This method provides a qualitatively correct description of three-dimensional materials, such as fulleride compounds. We consider a local Hamiltonian of the form $H_\text{loc}=\sum_\alpha Un_{\alpha\uparrow} n_{\alpha\downarrow}+\sum_{\alpha> \beta,\sigma}[U'n_{\alpha\sigma}n_{\beta\bar\sigma}+(U'-J)n_{\alpha\sigma}n_{\beta\sigma}]$, where $\alpha=1,2,3$ denotes the orbitals, $\sigma=\uparrow,\downarrow$ the spin, $U$ the intra-orbital repulsion, $U'$ the interorbital same-spin repulsion, and $J$ the Hund coupling. We set $U'=U-2J$ and $J=-U/4$, to be  consistent with the parameters used in the previous studies of electronic orderings \cite{Hoshino2017,Ishigaki2018,Ishigaki2019}. The density-density approximation enables an efficient solution of the DMFT equations \cite{ctqmc_prl2006,rmp_ctqmc}, and hence allows us to systematically compute the entropy of the model, as well as lattice susceptibilities, which require the measurement of four-point correlation functions \cite{Hoshino2015}.  

The infinitely connected Bethe lattice has a semi-circular density of states of bandwidth $W=4t$ (with $t$ a properly rescaled hopping), and in this study, we will use $W$ as the unit of energy. Magnetic order is suppressed by imposing the symmetry $G_{\alpha\uparrow}=G_{\alpha\downarrow}$ for the Green's functions, and we will search for solutions which satisfy $G_{1\sigma}=G_{2\sigma}$ (while the Green's function for orbital $3$ can be different). This is consistent with the symmetry breaking to the SOSM phase \cite{Hoshino2017} and with various types of orbital and charge order. 
The entropy per site $S$ is calculated with the procedure detailed in Ref.~\cite{Yue2020}, by first computing the total energy per site $E_\text{tot}$ of the system on a fine temperature grid and evaluating the specific heat $C_V(T)=dE_\text{tot}/dT$. Using the exactly known infinite-temperature result 
$S_\infty= -6\big[\frac{n}{6}\ln\frac{n}{6}+\big(1-\frac{n}{6}\big)\ln\big(1-\frac{n}{6}\big)\big]$
as a reference, we then compute $S(T)=S_\infty-\int_T^\infty dT' C_V(T')/T'$. From the total energy and entropy we also obtain the free energy $F=E_\text{tot}-TS$, which is useful for discussing the stability of different phases in a coexistence region \cite{Werner2007,PRL_2015_Haule_FreeEnergy}. 

Conventional electronic orders such as charge order (CDW), antiferro orbital order (AFO), ferro orbital order (FO) or superconductivity (SC), can be studied in two complementary ways. The first strategy, which will be used in Secs.~\ref{sec:doped_mott} and \ref{sec:doped_sosm} for the study of charge and orbital orders, is to break the symmetry explicitly in the DMFT calculation \cite{Chan2009}. An advantage of this approach is that it allows to discuss the stability regions of competing ordered phases. The second strategy, which will be used in Sec.~\ref{sec:susc}, is to compute lattice susceptibilities for the different orders using the local vertex from the DMFT solution in a Bethe-Salpeter equation, and to search for divergences in these susceptibilities \cite{Hoshino2015,Hoshino2017}. This approach allows to extract information on the tendencies towards different electronic orders from a single simulation.   

\section{Results}

\subsection{Entropy of the half-filled model with suppressed electronic orders}
\label{sec:entropy}

\begin{figure}[htp]
\includegraphics[clip,width=3.4in,angle=0]{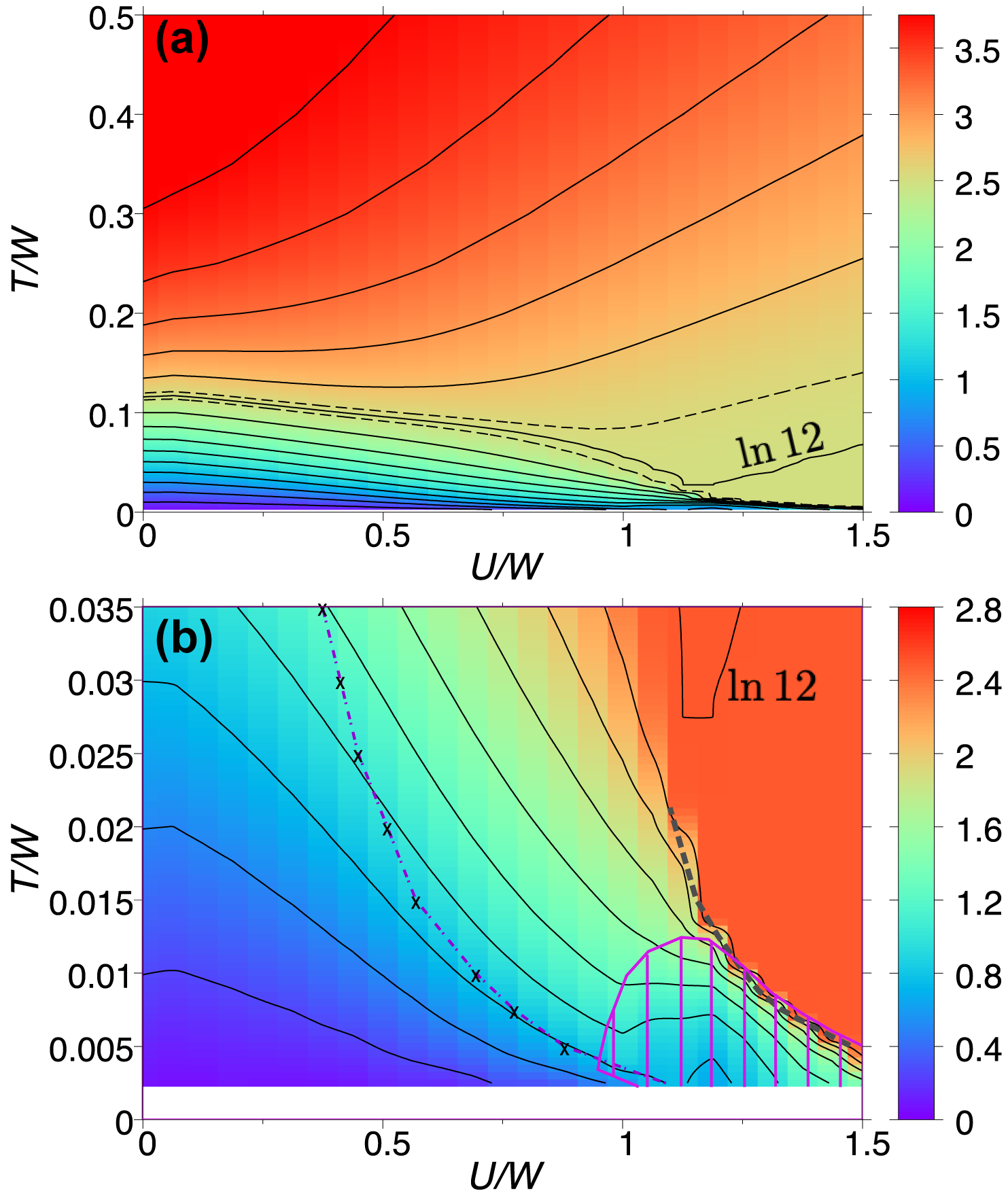} 
\caption{(color online). Contour maps of the entropy per site $S(U,T)$ in the plane of $U$ and $T$. 
(a) Large temperature range $T\in[0.0025:0.5]$. Here, the two dashed lines show the contours for $\ln 12 \pm 0.05$ to illustrate the extent of the Mott plateau. 
(b) Low temperature region $T\in[0.0025,0.035]$. The dash-dotted line indicates the locations of the maxima of $\Delta \chi^{\mathrm{orb}}_{\mathrm{loc}}$.
The dark-pink hashed region delimits the SOSM phase and the grey dashed line shows the Mott transition.
}
\label{fig:Entropy_TU_ntot3}
\end{figure}

We start by discussing the entropy per site $S$ of the half-filled model with suppressed antiferro-type orders by assuming, e.~g., the presence of geometrical frustration as in fcc lattices. Figure~\ref{fig:Entropy_TU_ntot3} plots the contour lines of the entropy in the interaction-temperature plane, with the top panel showing a wide temperature range, and the lower panel a zoom of the low-temperature region, with the SOSM solutions indicated by the pink hashing. In the SOSM phase the equivalence between the three orbitals is spontaneously broken \footnote{While Ref.~\cite{Ishigaki2019} reported two different types of SOSM phases, we will consider here only the phase with orbitals 1 and 2 in a paired Mott state, and the third orbital metallic.}. Also indicated as a gray dashed line is the Mott transition line which terminates at $T\simeq0.022$. We see that $S$ reaches approximately $\ln 12$ in the Mott phase, which is explained by the dominant configurations with one doubly occupied orbital, one empty orbital and one singly occupied orbital with spin up or down. The $S\approx \ln 12$ entropy plateau extends beyond the end point of the Mott transition line into the metal-insulator crossover region. Clearly separated from the Mott phase is the orbital-freezing crossover line (dashed line with crosses in the lower panel), which we define as the maximum in the dynamical contribution to the local orbital susceptibility, 
\begin{equation}
\Delta\chi_{\mathrm{loc}}^{\mathrm{orb}}=\int_{0}^{\beta}d\tau\left[\left\langle \tau^{\eta}_{i}(\tau)\tau^{\eta}_{i}(0)\right\rangle -\left\langle \tau^{\eta}_{i}(\beta/2)\tau^{\eta}_{i}(0)\right\rangle \right],
\end{equation}
where $\eta=8$, $\tau_{i}^{\eta}=\sum_{\gamma\gamma^{\prime}\sigma}c_{i\gamma\sigma}^{\dagger}\lambda_{\gamma\gamma^{\prime}}^{\eta}c_{i\gamma^{\prime}\sigma}$ and $\lambda^{\eta}$ is the $\eta$-th Gell-Mann matrix \cite{Hoshino2017}. This crossover line, which separates the Fermi liquid metal on the weak-$U$ side from an ``orbital frozen" metal on the large-$U$ side, has been shown in Ref.~\onlinecite{Hoshino2017} to correlate with the peak in the superconducting dome. We find that this line very roughly corresponds to the entropy contour for $S\approx 0.8 \gtrsim \ln 2$ at low temperatures. While the entropy in the orbital frozen metal is enhanced, the crossover is very broad. The transition into the SOSM phase is associated with a reduction in the entropy, compared to the orbital frozen metal. At low temperatures, the SOSM state exhibits an $\ln 2$ entropy per site, because there are two ways of distributing a doublon and a hole among the two paired Mott insulating orbitals. With increasing temperature, the entropy of the SOSM state increases, because of the metallic orbital $\alpha=3$, and also because the gap in the ``paired Mott" orbitals 1 and 2 starts to fill in; however, it increases less rapidly than in the orbital-frozen metal phase.  
 
\begin{figure}[t]
\includegraphics[clip,width=3.4in,angle=0]{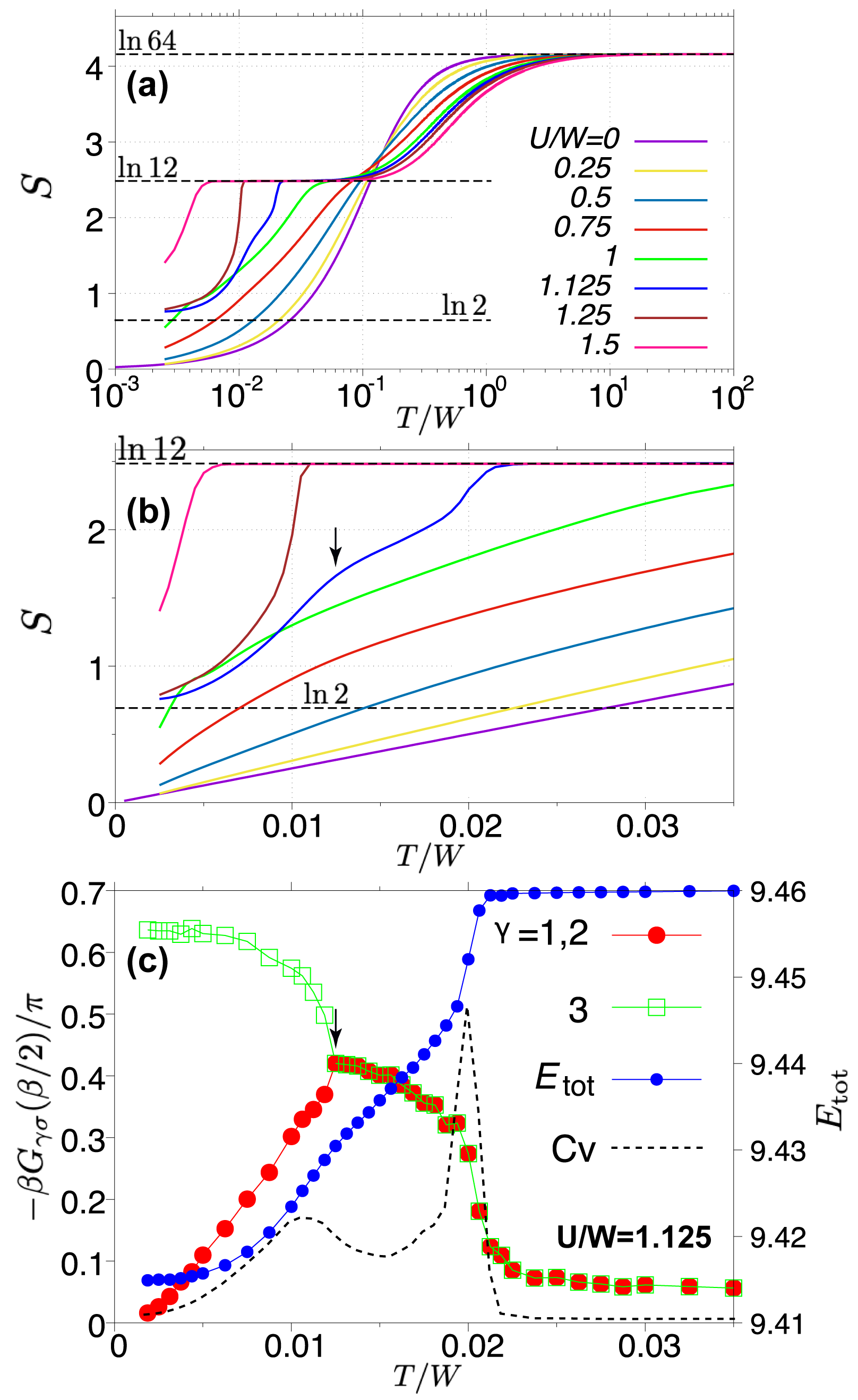} 
\caption{(color online). Entropy as a function of $T$ plotted on a logarithmic (a) and linear (b) temperature axis. 
Panel (c) shows $-\beta G(\beta/2)/\pi$ ($\approx$ DOS at the Fermi level), the total energy  $E_{\text{tot}}$ and 
the specific heat $C_V$ as a function of $T$ at $U=1.125$. The black arrow shows the temperature below which the SOSM phase emerges.
}
\label{fig:entropy_cuts}
\end{figure}

For a more detailed analysis, we show in Fig.~\ref{fig:entropy_cuts} $S(T)$ curves for fixed interactions. In panel (a), the results are plotted on a logarithmic scale, to illustrate how the $S_\infty=\ln 64$ value of the entropy is reached at very high temperatures in all cases, and how the entropy plateau with $S\approx \ln 12$ appears in and near the Mott phase. 
Panel (b) shows a zoom of the low-temperature region on a linear scale. Here, one can see that while the entropy in the metallic region goes to zero linearly as $T\rightarrow 0$, it is enhanced (with values larger than $\ln 2$) in the orbital-frozen regime. On the other hand, in the SOSM phase, it approaches $\ln 2$. The curve for $U=1.125$, which is on the metallic side of the Mott transition, exhibits several crossovers and a phase transition. As temperature is lowered below $T\approx 0.02$, there is a crossover from a bad metal state with suppressed quasi-particle peak and $S\approx \ln 12$ to a more coherent strongly correlated metal, and then, around $T\approx 0.014$ (black arrow) the transition into the SOSM phase, with a subsequent decrease of the entropy to $\ln 2$. The corresponding specific heat is plotted in panel (c), where we also show the quantity $-\beta G_{\gamma\sigma}(\beta/2)/\pi$ as a rough estimate of the density of states (DOS) near $\omega=0$ \cite{Gull2008}. Note that here all the orbitals are half-filled (no conventional orbital order) but the DOS can detect the broken orbital symmetry, reflecting a peculiarity of this ordered phase. While the DOS is suppressed in orbitals 1 and 2, compared to orbital 3, the Mott insulating nature is evident only at low temperatures. Near $T\approx 0.014$, the SOSM state is in fact a metallic state with a spontaneous symmetry breaking into two ``bad metallic" and one ``good metallic" orbital. 

At $U=1.0$, the system just barely crosses into the SOSM phase as $T$ decreases. It undergoes crossovers and transitions from bad metal to good metal, to
SOSM, and (below $T\approx 0.004$) makes a  transition to the low-temperature Fermi liquid. As shown in the Appendix (Fig.~\ref{fig:Aw0_Cv_Etot_UxW}(a)), there are three peaks in $C_V$ associated with these  
transitions or crossovers. The first two peaks are broad and overlap with each other. The peak associated with the transition from SOSM phase to the good metal below $T\approx 0.004$ is very sharp (in reality a delta function, if the transition is first order), and responsible for the decrease of the entropy below $\ln 2$, 
as shown by the green line in Fig.~\ref{fig:entropy_cuts}(a-b). The results for $U=1.25$, on the Mott insulating side, are shown in Fig.~\ref{fig:Aw0_Cv_Etot_UxW}(b), where the $C_V$ peak associated with the transition into the SOSM phase also exhibits a delta-function like contribution, but extends to low temperatures due to the strongly correlated nature of the metallic orbital.

\begin{figure}[t]
\includegraphics[clip,width=3.4in,angle=0]{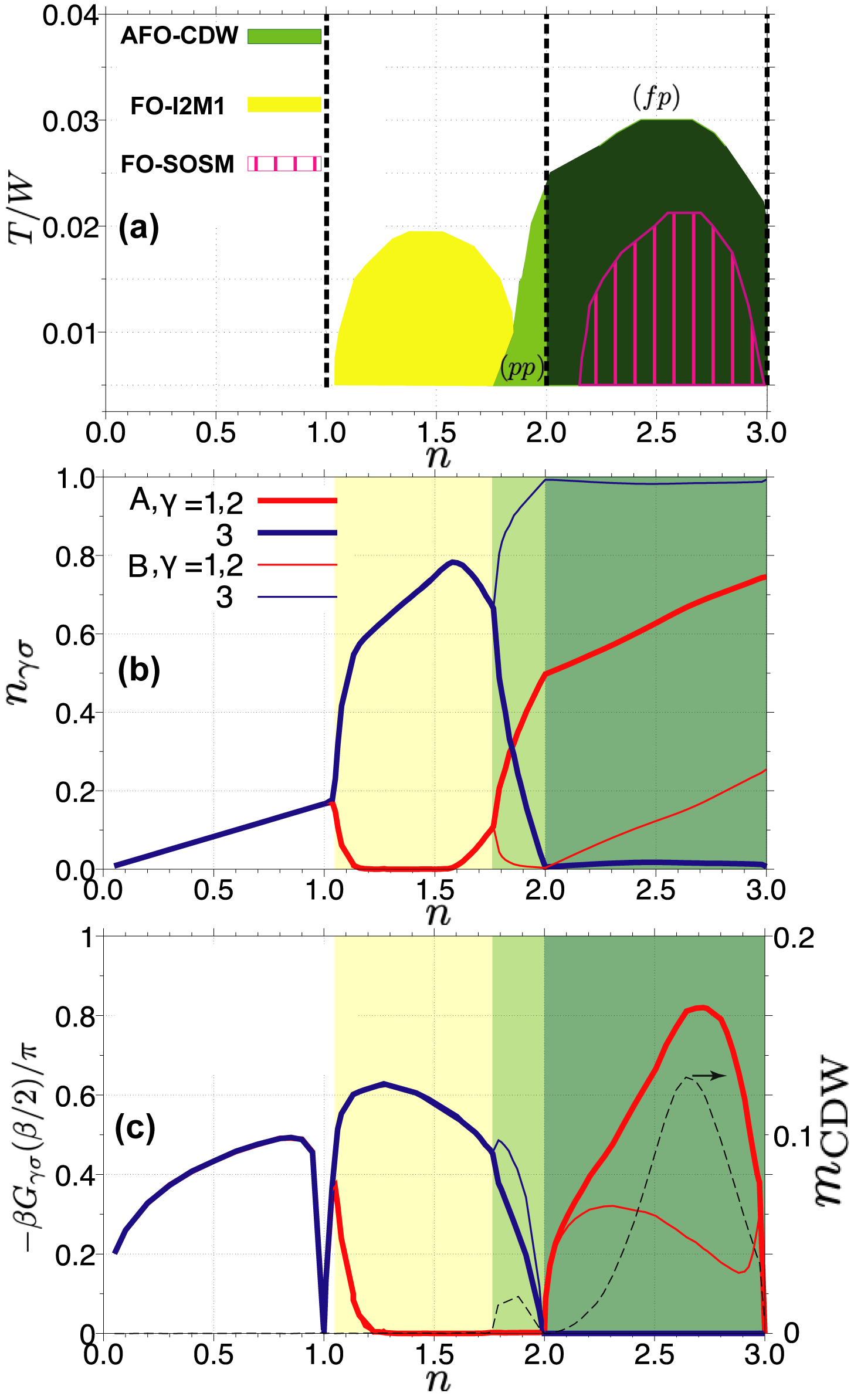}
\caption{(color online). (a) Phase diagram at $U=1.5$ in the space of filling and temperature. The stable phases are the metal (white region), Mott insulator (black dashed), 
AFO$^{(\text{fp})}$-CDW (dark green), AFO$^{(\text{pp})}$-CDW (light green) and FO-I2M1 (yellow). The FO-SOSM phase (pink hashed) is metastable. 
(b) Filling per spin-orbital for the stable phases as a function of average filling at $T=0.005$. 
(c) DOS at the Fermi level for $T=0.005$. 
The dashed line in panel (c) shows the filling dependence of the CDW order parameter $m_{\text{CDW}}=|n_A-n_B|$.
(The spectral functions for the stable phases at $n=1.5$, $1.79$ and $2.5$ are shown in Fig.~\ref{fig:Aw_T0p01_U3D_n}.)
}
\label{fig:phase_diagram_doping}
\end{figure}

\subsection{Electronic orders in the doped Mott regime}
\label{sec:doped_mott}

The phase diagram at $U=1.5$ in the space of  filling $n\in[0,3]$ and temperature is shown in Fig.~\ref{fig:phase_diagram_doping}. 
Now we allow for translational symmetry breaking and consider antiferro-type orders, as relevant e.g. in the case of bcc lattices. Since we suppress magnetic orders, the half-filled Mott insulator is at low temperatures in a fully polarized AFO state, which we call AFO$^{(\text{fp})}$. In this state, orbital 3 is almost empty on sublattice A and almost fully occupied on sublattice B, while orbitals 1 and 2 are 3/4 filled on sublattice A and 1/4 filled on sublattice B. 
As we hole-dope this state, the orbitals 3 remain fully polarized (band insulating), while the occupation in orbitals 1 and 2 is reduced proportional to the doping on both sublattices. The spectral functions in Fig.~\ref{fig:Aw_T0p01_U3D_n}(c) show that the partially filled orbitals in the doped AFO$^{(\text{fp})}$ state are metallic. 
At $n=2$ the occupations in the three orbitals reach $n_{\gamma\sigma}=(0.5,0.5,0)$ on sublattice A and $(0,0,1)$ on sublattice B, and the half-filled orbitals undergo a Mott transition. 
If we further hole dope this $n=2$ insulator, we obtain a partially polarized AFO state, denoted by AFO$^{(\text{pp})}$, where the orbital differentiation between sublattice A and B rapidly shrinks with hole doping, resulting in a state with large occupation in orbital 3 and small occupation in orbitals 1 and 2, and a transition to a state with FO order. The AFO solutions also exhibit a small CDW with a maximum $|n_A-n_B|\approx 0.13$ reached near $n\approx 2.65$ at $T=0.005$, as indicated by the dashed line in Fig.~\ref{fig:phase_diagram_doping}(c). Hence, we denote the green phases in Fig.~\ref{fig:phase_diagram_doping} by AFO$^{(\text{fp})}$+CDW and AFO$^{(\text{pp})}$+CDW.

The FO state with large occupation in the metallic orbital 3 and low occupations in the orbitals 1 and 2 persists down to filling $n=1$, where the system becomes again Mott insulating. In most of this filling range, orbital 1 and 2 are in fact empty (band insulating, see spectral functions in  Fig.~\ref{fig:Aw_T0p01_U3D_n}(a)), which is why we denote this phase as FO-I2M1 (two insulating orbitals and one metallic orbital). 
For fillings below 1 there is no orbital or charge order. 

There exist also metastable electronically ordered phases, which can be stabilized in DMFT by starting from an appropriate initial solution.
One of these metastable phases is the doped version of the SOSM phase, with one metallic and two paired Mott insulating orbitals. Away from half-filling, this phase (which is indicated by the pink hashing in Fig.~\ref{fig:phase_diagram_doping}(a)) has FO order, because the Mott insulating orbitals remain essentially half-filled, while the metallic one is doped. We call this the FO-SOSM phase, and show the corresponding orbital fillings in Fig.~\ref{fig:metalstable_phase_orbital_filling_doping} in Appendix~\ref{app_metastable}. This phase is continuously connected to the SOSM phase appearing next to the Mott insulator in the half-filled phase diagram (pink hashed region in Fig.~\ref{fig:Entropy_TU_ntot3}). An interesting observation is that the FO-SOSM phase exists up to higher temperatures ($T_\text{max} > 0.022$ at $U=1.5$, $n=2.6$) than the half-filled SOSM state ($T_\text{max} 
 \approx 0.013$ at $U=1.125$, $n=3$). Hence, if the AFO$^{(\text{fp})}$+CDW order can be suppressed by geometrical frustration, e. g. in fulleride compounds with an fcc lattice, then electron or hole doping of the Mott insulating compounds should result in a FO-SOSM state which is stable over a wide doping range and up to high temperatures.  

\begin{figure}[t]
\includegraphics[clip,width=3.4in,angle=0]{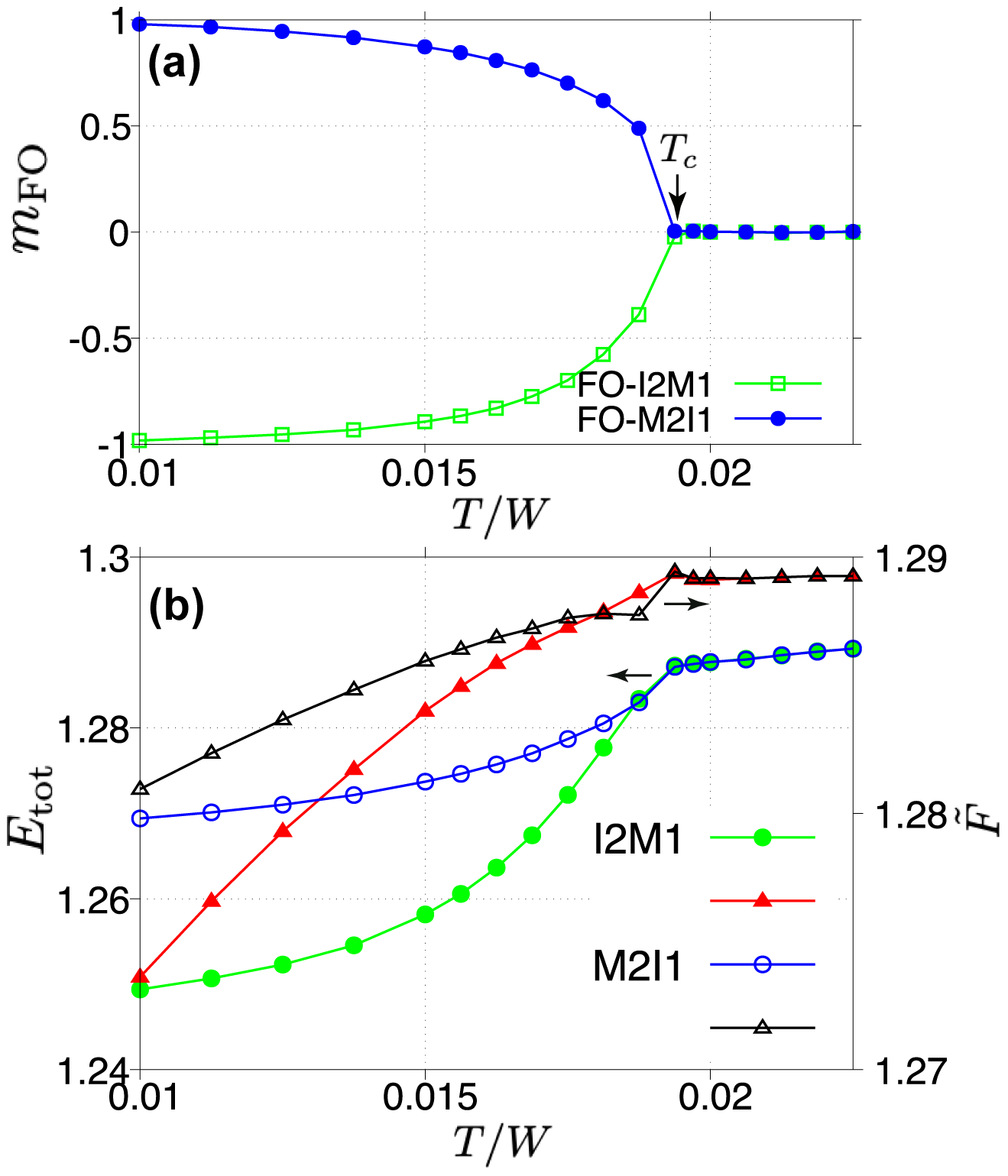}
\caption{(color online). Temperature dependence of the total energy $E_\text{tot}$ and relative free energy $\tilde F(T)=F(T)-F(T_c)$ near the phase transition 
from the metal to the FO phases at filling $n=1.5$. Panel (a): order parameter $m_{FO}=\frac{(n_1+n_2)/2-n_3}{(n_1+n_2)/2+n_3}$ for FO-I2M1 and FO-M2I1. 
The black arrow indicates that both FO orders appear below the same temperature $T_c$. Panel (b): $E_\text{tot}(T)$ and $\tilde F(T)$ for both FO phases.
}
\label{fig:mFO_free_energy_ntot1p5}
\end{figure}

In the filling range $1<n\lesssim 2$ we also find a second FO solution, with two metallic and one (band) insulating orbital, as also indicated in Fig.~\ref{fig:metalstable_phase_orbital_filling_doping}. We call this the FO-M2I1 phase (two metallic orbitals and one insulating orbital as shown in Fig.~\ref{fig:Aw_T0p01_U3D_n}(b) where the $\gamma=3$ orbital is empty). 
This phase can be stabilized in almost the same parameter region as the FO-I2M1 phase, but it has a larger energy $E_\text{tot}$ and free energy $F$ in most of this region (except close to $T_c$). To illustrate this, we plot in Fig.~\ref{fig:mFO_free_energy_ntot1p5} the total energies and the free energies relative to the critical temperature $T_c\approx0.019$, $\tilde F(T)=F(T)-F(T_c)$, at $n=1.5$ as a function of $T$.  These data show that below $T\approx 0.018$, the stable phase is FO-I2M1.

There is also a metastable phase overlapping with the AFO$^{(\text{pp})}$-CDW phase, which connects to the  FO-M2I1 phase and can be stabilized for example by increasing the filling starting from this FO phase, but not by decreasing the filling from $n=2$. We call this the AFO$^{(\text{pp2})}$-CDW phase and show the corresponding orbital-dependent fillings in Fig.~\ref{fig:metalstable_phase_orbital_filling_doping}.

\subsection{Electronic orders in the doped SOSM regime}
\label{sec:doped_sosm}

\begin{figure}[t]
\includegraphics[clip,width=3.4in,angle=0]{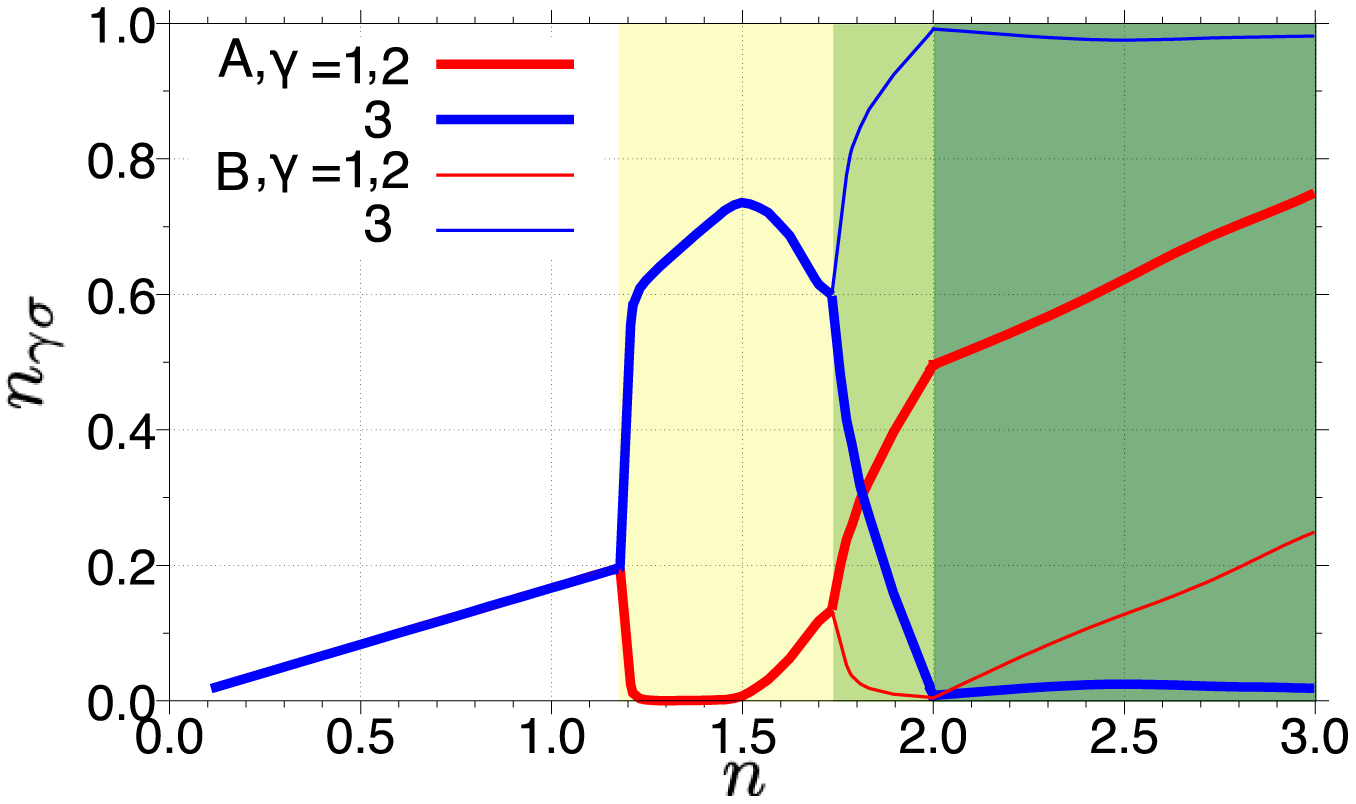}
\caption{(color online). Filling per spin-orbital for the stable phases at $U=1.25$, $T=0.005$.
The meaning of the colored background is the same as in Fig.~\ref{fig:phase_diagram_doping}. 
}
\label{fig:occ_SOSM1_dope_U2p5T0p01}
\end{figure}

We next consider the filling dependence of the electronic orders at $U=1.25$, where the half-filled system below $T\approx 0.01$ is in the SOSM phase (in the absence of conventional electronic orders), or in the AFO$^{(\text{fp})}$+CDW phase. Here we do not map out the entire phase diagram in the filling-temperature plane but merely discuss the cut at $T=0.005$. The occupations of the spin-orbitals are plotted as a function of total filling in Fig.~\ref{fig:occ_SOSM1_dope_U2p5T0p01} (stable phases) and Fig.~\ref{fig:metalstable_phase_orbital_filling_doping}(b) (metastable phases). The results are similar to those shown for $U=1.5$ in Fig.~\ref{fig:phase_diagram_doping}, but the FO-I2M1 phase is stable only down to filling $n\approx 1.2$, while the filling range of the AFO$^{(\text{pp})}$+CDW phase slightly expands. Also, the metastable FO-SOSM phase directly connects to the (also metastable) half-filled SOSM phase, while in the doped Mott case ($U=1.5$, $T=0.005$, Fig.~\ref{fig:metalstable_phase_orbital_filling_doping}(a)) the FO-SOSM state exists only above some critical doping concentration.
 In both cases, the FO-SOSM state can be stabilized over a wide doping range, down to $n\approx 2.2$. The orbital differentiation in $n_{\gamma\sigma}$ vanishes in the limit $n\rightarrow 3$ as shown in Fig.~\ref{fig:metalstable_phase_orbital_filling_doping}(b) but the orbital symmetry breaking still remains there, which distinguishes the SOSM state from conventional FO orders. Based on these results, and the data in Fig.~\ref{fig:Entropy_TU_ntot3}, we identify in Fig.~\ref{fig:phase_diagram1} the FO phase at $n>2$, which is detected by the susceptibility analysis, with the \mbox{(FO-)SOSM} phase. 

\subsection{Orbital freezing and superconductivity in the doped system}
\label{sec:susc}

In this section we complement the previous results on electronic ordering instabilities with data obtained from susceptibility calculations. With this approach, we can also investigate the $s$-wave superconducting instability, without introducing Nambu Green's functions in the DMFT simulations. This allows us to address the connection between orbital freezing and superconductivity in the doped system. The half filled model with $J=-U/4$ has been analyzed with the same  technique in Ref.~\onlinecite{Hoshino2017}. 

\begin{figure}[t]
\includegraphics[clip,width=3.4in,angle=0]{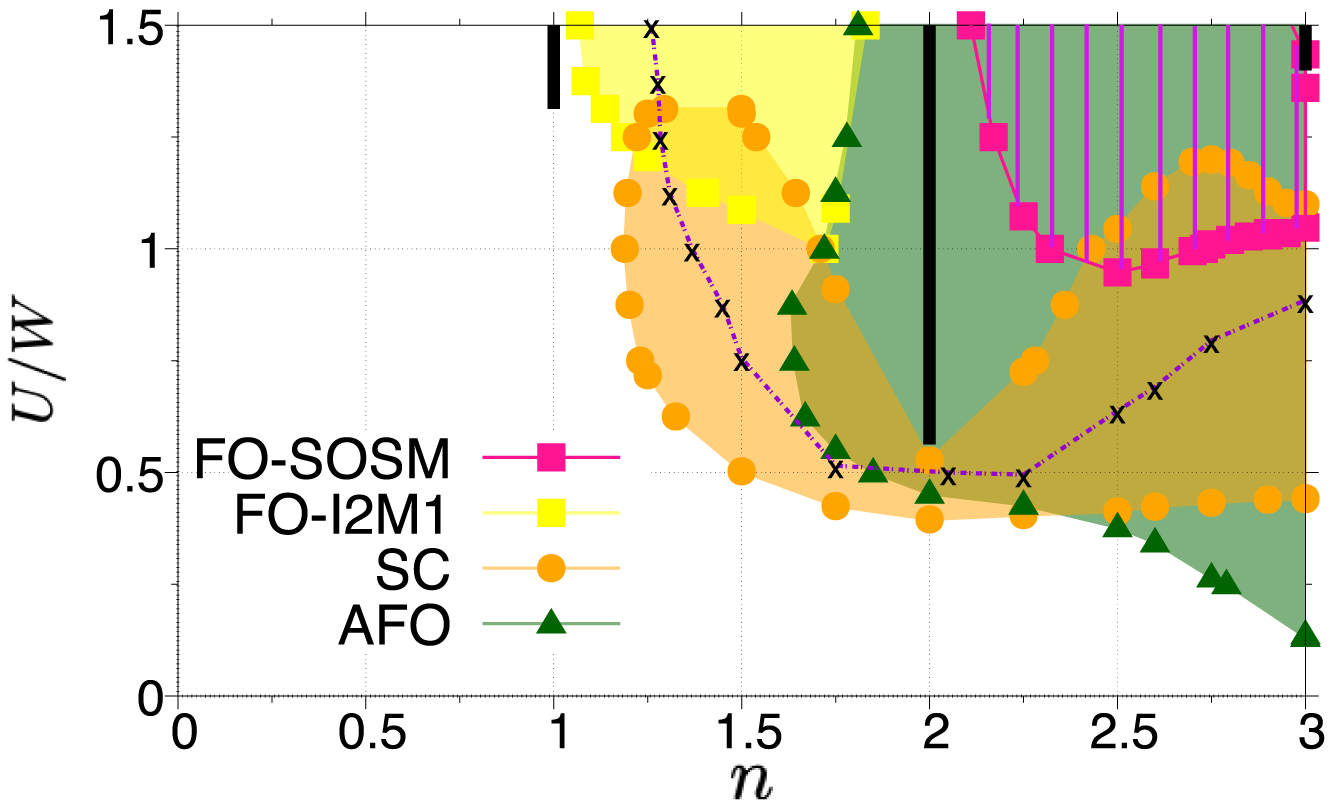}
\caption{(color online). 
Stability regions of conventional electronic orders at $T=0.005$, determined by the divergence in the corresponding lattice susceptibilities.
The dash-dotted line indicates the locations of the maxima of $\Delta \chi^{\mathrm{orb}}_{\mathrm{loc}}$.
}
\label{fig:phase_diagram1}
\end{figure}

We start by plotting in Fig.~\ref{fig:phase_diagram1} the potential stability regions of different phases in the space of $n$ and $U$ at temperature $T=0.005$. We see that AFO order 
(in this analysis we do not distinguish between AFO$^{(\text{fp})}$ and AFO$^{(\text{pp})}$)
covers a wide $U$ region in the filling range $2\le n \le 3$ and also for some range of fillings below $n=2$, while FO order potentially appears at large $U$ for fillings $n>1$, except in the vicinity of $n=2$. As mentioned in Sec.~\ref{sec:model}, the susceptibility analysis cannot tell us which of two overlapping orders is stable, but it is natural to assume that the order which covers the larger parameter region is actually realized. Within the accuracy of our analysis, the results are then consistent with the data presented in the previous sections, which for $U=1.5$ demonstrated AFO order down to $n\approx 1.76$ (1.81 in Fig.~\ref{fig:phase_diagram1}) at $T=0.005$ followed by FO order which is stable down to $n\approx 1.05$. Similarly, at $U=1.25$ we found AFO order down to $n\approx 1.73$  (1.78 in Fig.~\ref{fig:phase_diagram1}) and FO order down to $n\approx 1.2$.

We also indicate by orange dots the potential stability region of the SC phase, and by the violet line with black crosses the orbital freezing crossover (maxima of $\Delta\chi_\text{loc}^\text{orb}$). At half-filling, the SC phase appears detached from the Mott phase next to the SOSM phase. (This is different from the two-orbital model, where the paired Mott state is very stable \cite{Steiner2016}, and the SOSM phase does not exist.) Upon hole doping, the SC phase expands to larger $U$, 
before it shrinks considerably as we approach the very stable $n=2$ Mott insulator. For fillings below $n=2$ the SC region expands again and reaches its largest extent in the $U$ direction near $n=1.5$. 

\begin{figure}[t]
\includegraphics[clip,width=3.4in,angle=0]{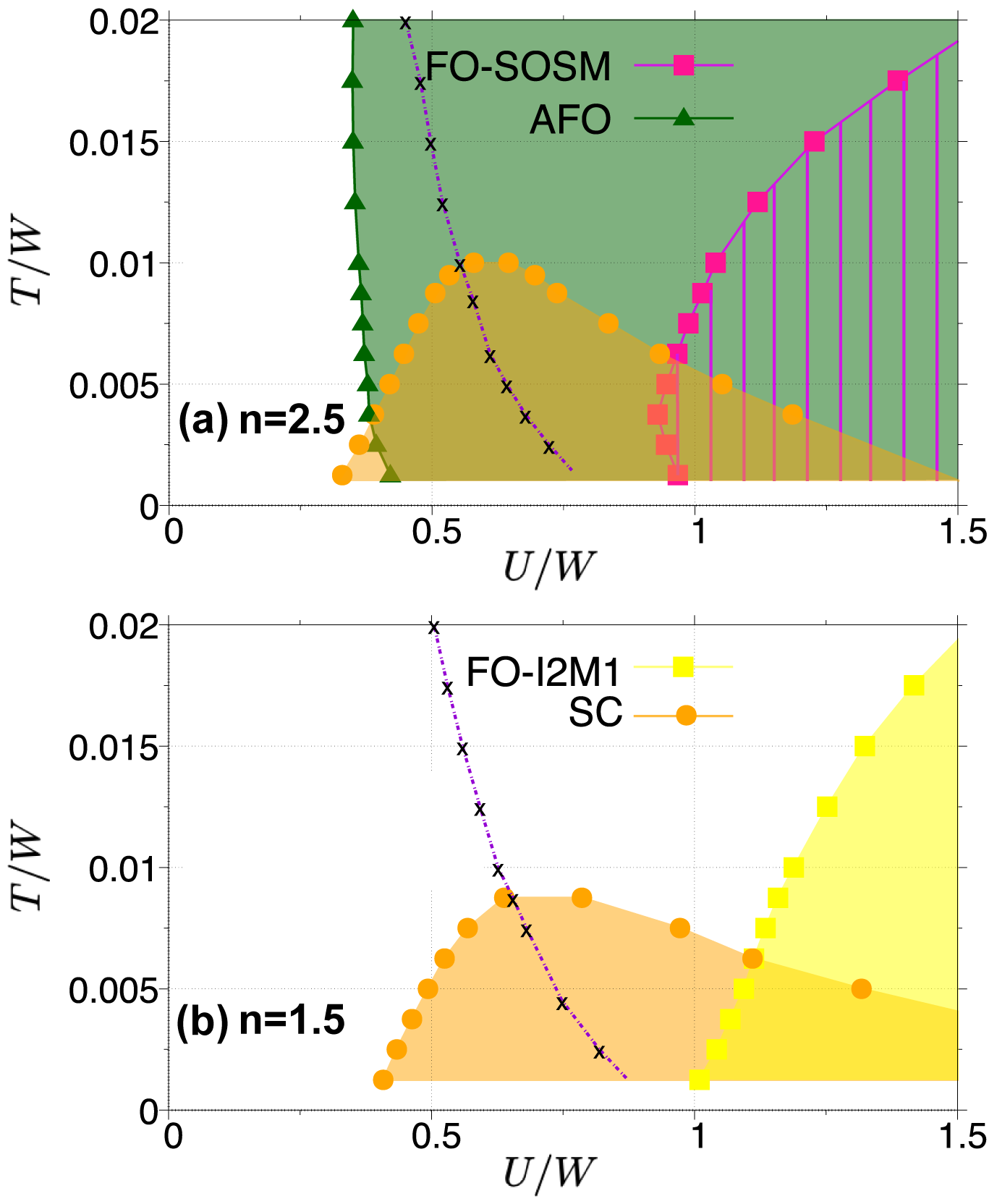}
\caption{(color online). 
Phase diagrams in the $U$-$T$ plane determined by the susceptibility calculations for $n=1.5$ (a) and $n=2.5$ (b).
The dash-dotted line indicates the locations of the maxima of $\Delta \chi^{\mathrm{orb}}_{\mathrm{loc}}$.
}
\label{fig:phase_diagram2}
\end{figure}

Independent of filling, the highest superconducting $T_c$ is reached near the orbital freezing crossover. This is suggested already by the obvious connection between the orbital freezing line in Fig.~\ref{fig:phase_diagram1} and the shape of the SC region, and demonstrated explicitly in Fig.~\ref{fig:phase_diagram2}, which shows phase diagrams in the $U$-$T$ plane for $n=2.5$ and $1.5$, respectively. The peaks of the SC domes appear in 
 the orbital freezing crossover region, which demonstrates that the unconventional SC state in $J<0$ multi-orbital systems is induced by the emergence of slowly fluctuating orbital moments, in analogy to the spin-freezing induced superconductivity in $J>0$ multi-orbital systems \cite{Hoshino2015,Steiner2016,Werner2016}.  

While SC near $n=2.5$ can only be realized by suppressing AFO order, or on the small-$U$ side of the dome at very low $T$ (see Fig.~\ref{fig:phase_diagram2}(a)), there are no competing electronic orders in the spin freezing crossover region near $n=1.5$. The maximum $T_c$ at $n=1.5$ and $2.5$ is comparable to the maximum $T_c\approx 0.01$ reached at half-filling \cite{Hoshino2015}. 
In fcc fulleride superconductors, AFO is suppressed by geometrical frustration, and SC has been recently found also in doped compounds \cite{Ren2020,Han2020}, although a more careful modeling would be needed to account for the thin-film nature of these systems. Our results suggest that also the quarter-filled fulleride compounds should exhibit unconventional SC with a high $T_c$.

\section{
Discussion and Summary
}
\label{sec:summary}

We have studied the filling dependence of electronically ordered states in the three-orbital Hubbard model with negative Hund coupling and degenerate bands. This model is relevant for the description of fulleride compound, where the small bare $J$ is overscreend by a coupling to Jahn-Teller phonons \cite{Erio_prb1997_JahnTeller,RevModPhys.81.943,Nomura2012}. Recent experimental progress \cite{Ren2020,Han2020} enables a systematic study of chemically doped thin films of these unconventional superconductors. Our work is not meant to be a realistic study of fulleride compounds, but the simple model considered captures the qualitative physics of this class of materials, as has been shown in recent investigations of the half-filled system \cite{Hoshino2017,Ishigaki2018,Ishigaki2019}. In particular, these theoretical studies have explained the puzzling ``Jahn-Teller metal" phase \cite{Zadik2015} as a manifestation of the SOSM phase. This composite ordered phase is unique to multi-orbital Hubbard systems with negative Hund coupling, and conceptually related to odd-frequency order in the two-channel Kondo problem \cite{Hoshino2014,Hoshino2019}. The previous investigations of the half-filled model have also clearly revealed that the unconventional superconductivity is induced by an orbital freezing crossover \cite{Hoshino2017}.

In this work, we have focused on orbital orders, the SOSM phase, and superconductivity, and investigated their stability as the system is doped away from half-filling. (Antiferromagnetic order also exists in this model, but it appears only in and near the $n=1$ and $n=3$ Mott insulating phases.) At low temperatures, the phase diagram for fillings $n\gtrsim 1.75$ is dominated by antiferro orbital order. At large interactions $U$,  
FO order appears for $1<n\lesssim 2$. If the AFO state is suppressed, e.~g. by geometric frustration as in fcc fullerides, superconductivity appears in a wide $U$ range between $2\le n\le 3$, and it reaches maximum $T_c$s which are comparable to the undoped system. Also for $1.25 \lesssim n<2$ superconductivity is prominent and reaches similarly high $T_c$s near $n=1.5$. This filling regime is particularly interesting, because here the superconducting state for $U<2$ does not compete with other electronic instabilities. 

Also away from half-filling, the highest superconducting $T_c$s are reached in the orbital-freezing crossover region, where slowly fluctuating orbital moments emerge within the metal phase. This connection to orbital freezing motivated us to also study the entropy of the three-orbital model with $J<0$. Focusing on the half-filled model for simplicity, we indeed found the expected enhancement of the entropy in the orbital frozen metal regime, although the crossover is rather broad. This is consistent with enhanced orbital fluctuations in a wide interaction range and a correspondingly wide superconducting dome.   
A more prominent feature of the entropy is a $\ln 12$ plateau in the half-filled Mott state, which extends to high temperatures, and deep into the metal-insulator crossover region. We also showed that the entropy of the SOSM state is lower than that of the orbital-frozen metal. 

While the SOSM phase is dominated by the AFO phase in our model, it can exist as a stable phase in geometrically frustrated systems \cite{Zadik2015}. It is thus interesting to ask how this composite ordered state evolves under doping. While the half-filled SOSM state at higher temperatures is in fact metallic (with two bad metallic and one good metallic orbital), it is characterized at low temperatures by the spontaneous symmetry breaking into two paired Mott insulating orbitals and one metallic orbital. Upon hole doping, the Mott insulating orbitals remain half-filled (and insulating), while the filling of the metallic orbital is reduced. The doped SOSM state thus exhibits a conventional FO order. 

Not only does the half-filled SOSM state evolve under hole doping into a FO-SOSM state which is stable over a wide doping range, a similar FO-SOSM state can also be induced at larger interactions, by hole-doping the half-filled Mott insulator. In this doped large-$U$ region, the $T_c$ of the FO-SOSM phase can be higher than the maximum $T_c$ of the half-filled SOSM (Jahn-Teller metal) phase. In experiments on doped Mott insulating fulleride compounds, e.~g. doped Cs$_3$C$_{60}$ \cite{Tc_38K_Cs3C60,Zadik2015}, one may thus find a prominent FO-SOSM phase, with experimental signatures similar to those of the Jahn-Teller metal.

\acknowledgments{
The calculations have been performed on the Beo05 cluster at the University of Fribourg, 
using a code based on iQist \cite{HUANG2015140,iqist}, and at the facilities of
the Supercomputer Center of the Institute for Solid State Physics, University of Tokyo. 
We acknowledge support by SNSF Grant No. 200021-165539 and JSPS KAKENHI Grant No.~JP18K13490. 
}

\clearpage

\appendix
\numberwithin{equation}{section}
\numberwithin{figure}{section}

\begin{widetext}

\section{Specific heat and orbital occupations near the SOSM phase}

In Fig.~\ref{fig:Aw0_Cv_Etot_UxW} we plot the total energy, specific heat, and orbital occupations as a function of temperature for $U=1$ (panel (a)) and 1.25 (panel (b)). In the $U=1$ case, the system enters the SOSM phase around $T=0.013$ and switches to a  strongly correlated Fermi liquid around $T=0.004$, which is accompanied by a large increase in the specific heat. (Since the transition is expected to be first order, the specific heat exhibits a delta-function peak at the transition point.) At $U=1.25$ the system makes a transition from the Mott insulator to the SOSM phase at $T\approx 0.01$, again with a delta-function like peak in the specific heat at the transition point and an enhanced specific heat in the correlated Fermi liquid state at $T>0$. 

\begin{figure*}[htp]
\includegraphics[clip,width=0.835\paperwidth,angle=0]{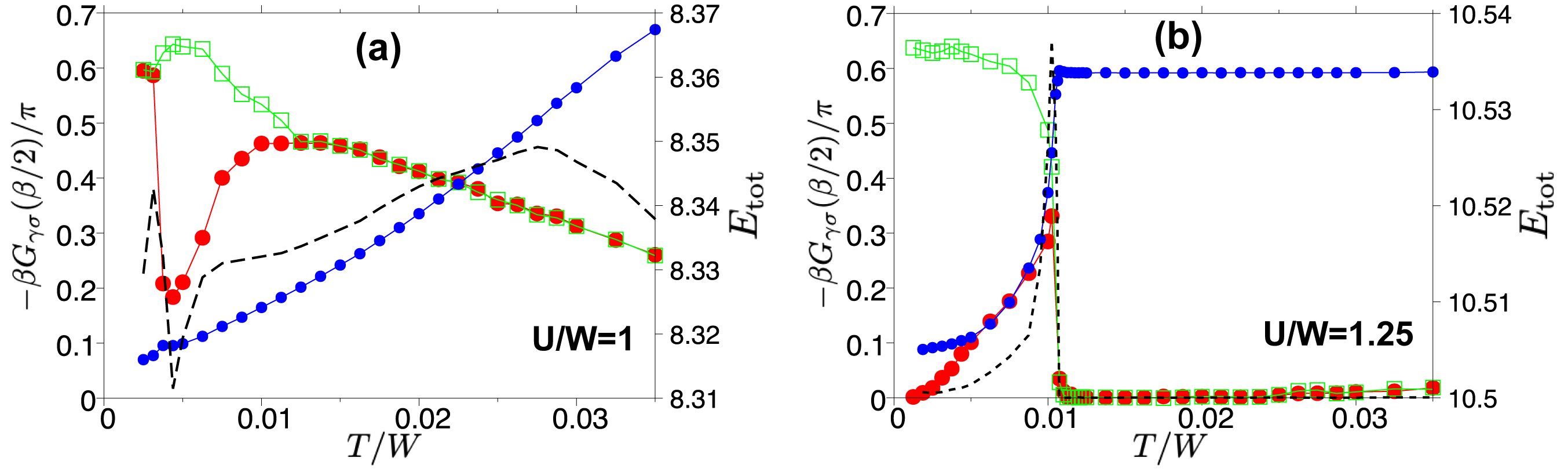}
\caption{(color online). Supplementary data to Fig.~\ref{fig:entropy_cuts}. 
Green and red points show $-\beta G(\beta/2)/\pi$ ($\approx$ DOS at Fermi level) for orbitals $1$, $2$ and $3$, respectively, the blue points the total energy  $E_{\text{tot}}$ and the black dashed line the
specific heat $C_V$ at (a) $U=1$ and (b) $U=1.25$.
}
\label{fig:Aw0_Cv_Etot_UxW}
\end{figure*}

\section{Metastable electronically ordered phases}
\label{app_metastable}

We illustrate in Fig.~\ref{fig:metalstable_phase_orbital_filling_doping} the stability regions and orbital occupations of the metastable phases, and the transitions induced by changing the chemical potential $\mu$ in these metastable phases. 
The green (black) thick arrows show the phase transitions observed by increasing (decreasing) the chemical potential $\mu$. 
By decreasing $\mu$ from the Mott phase at $n=3$, $U=1.5$, the system ends up in the AFO$^{(\text{fp})}$-CDW phase, as indicated by the $\nwarrow$ arrow in panel (a). The stability range of the AFO$^{(\text{fp})}$-CDW phase is very narrow because $T=0.005$ is very close to the transition between the half-filled Mott and SOSM phases. At higher temperatures, this phase will extend over a wider doping range, while at lower temperatures, it would disappear completely. 

By decreasing $\mu$ from the SOSM phase at $n=3$, $U=1.25$, an orbital polarization is induced, and the system makes a transition into the FO-SOSM phase, as indicated by the $\nwarrow$ arrow in panel (b). 

\begin{figure*}[htp]
\includegraphics[clip,width=0.835\paperwidth,angle=0]{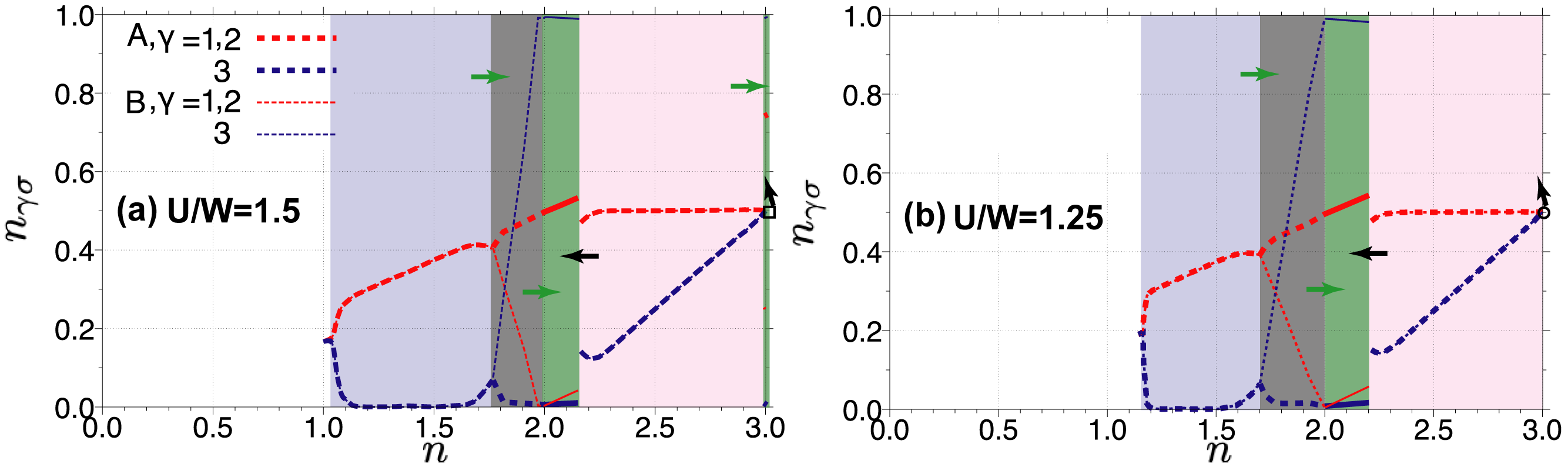}
\caption{(color online). 
Filling per spin-orbital in the metastable phases and phase transitions from metastable phases to stable ones induced by changing $\mu$. In both panels $T=0.005$. Panel (a) shows results for $U=1.5$ and panel (b) for $U=1.25$. The metastable phases are the FO-M2I1 phase (light blue), the FO-SOSM phase (light pink) and 
the AFO$^{(\text{pp2})}$-CDW phase (black hashed). The dark green region indicates the stable AFO$^{(\text{fp})}$-CDW phase (see Fig.~\ref{fig:phase_diagram_doping}(a) for the full stability region of this phase).
The black box in panel (a) denotes the Mott phase at $n=3$, $U=1.5$, while the black circle in panel (b) denotes the SOSM phase at $n=3$, $U=1.25$.
}
\label{fig:metalstable_phase_orbital_filling_doping}
\end{figure*}

\section{Spectral functions}

Figure~\ref{fig:Aw_T0p01_U3D_n} plots representative spectral functions for the three stable orbitally ordered phases at $T=0.005$. 
Panel (a) shows an example for the FO-I2M1 phase at $n=1.5$, with orbital 3 metallic and orbitals 1 and 2 band insulating (empty). 
In contrast, panel (b) shows an example for the FO-M2I1 phase, with orbital 1 and 2 metallic and orbital 3 band insulating (empty). 
Panel (c) is an example for the AFO$^{(\text{pp})}$-CDW state at $n=1.79$. Here, we have different solutions for the two sublattices, with orbital 3 metallic and orbitals 1 and 2 essentially empty. Panel (d) plots the spectral functions at $n=2.5$, in the AFO$^{(\text{fp})}$-CDW phase. Here, orbital 3 is band insulating (full or empty) while orbitals 1 and 2 are metallic. 

\begin{figure*}[htp]
\includegraphics[clip,width=0.835\paperwidth,angle=0]{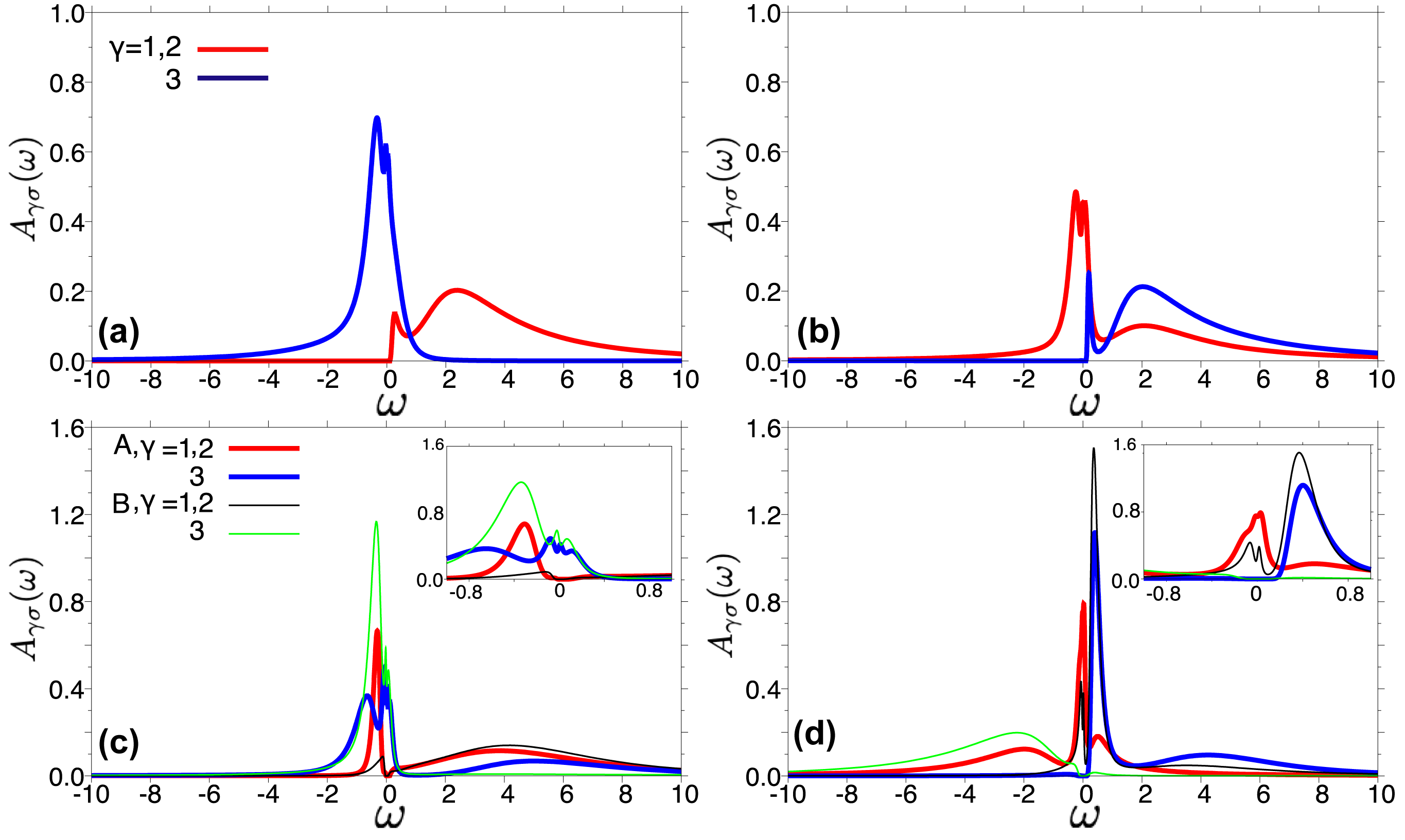}
\caption{(color online). Spectral functions at $T=0.005$, where panel (a) corresponds to the FO-I2M1 phase at $n=1.5$, 
panel (b) to the FO-M2I1 phase at $n=1.5$, panel (c) to the AFO$^{(\text{pp})}$-CDW phase at $n=1.79$ and 
panel (d) to the AFO$^{(\text{fp})}$-CDW phase at $n=2.5$.  The insets in panels (c) and (d) show $A(\omega)$ in the range $ -1<\omega <1$.
}
\label{fig:Aw_T0p01_U3D_n}
\end{figure*}
\end{widetext}

\clearpage
\bibliography{Entropyfullerides.bib}

\end{document}